\documentclass[epj]{svjour}
\usepackage{graphicx}

\usepackage{color}

\newcommand{\ie}{\emph{i.e.\/}} 

\newcommand{\etal}{\emph{et al\/}}
\newcommand{\zeff}{\ensuremath{Z_\mathrm{eff}}}
\newcommand{\zi}{\ensuremath{Z_\mathrm{int}}}
\newcommand{\Ang}{\,\mathring \mathrm{A}}
\newcommand{\csalt}{\ensuremath{c_\mathrm{salt}}}

\begin{document}
\title{Simulation of charge reversal in salty environments: Giant overcharging?}
\author{Olaf~Lenz\inst{1} \and Christian Holm\inst{1,2}
}                     
\mail{olenz@fias.uni-frankfurt.de, holm@fias.uni-frankfurt.de}
\institute{Frankfurt Institute for Advanced Studies,
  Johann-Wolfgang-Goethe-Universit\"at, Frankfurt am Main, Germany \and
  Max-Planck-Institut f\"ur Polymerforschung, Mainz, Germany}
\date{Received: date / Revised version: date}
%
\abstract{ We have performed MD simulations of a highly charged
  colloid in a solution of 3:1 and additional 1:1 salt.  The
  dependency of the colloid's inverted charge on the concentration of
  the additional 1:1 salt has been studied.  Most theories predict,
  that the inverted charge increases when the concentration of
  monovalent salt grows, up to what is called \emph{giant
    overcharging}, while experiments and simulational studies observe
  the opposite.  Our simulations agree with the experimental findings
  and shed light onto the weaknesses of the theories.
} 
\maketitle
\section{Introduction}
\label{intro}

In this work we study the phenomenon of overcharging on the basis of
the simple restrictive primitive model.  When a highly charged
colloid (or \emph{macroion}) is put into a solution that contains
multivalent counterions, its charge can become overcompensated, such
that the effective charge of the colloid-ion-complex becomes
oppositely charged. In the literature, this phenomenon is called
\emph{charge reversal}, \emph{charge inversion} or
\emph{overcharging}.

Experimentally, the phenomenon has been demonstrated by
electrophoresis experiments, where a reversed mobility has been
observed for charged colloids in a solution that contained multivalent
ions (for a review, see \cite{quesada-perez03a}).  Only recently,
experiments were able to directly measure charge reversal in a system
of a solution of multivalent salt at a charged wall
\cite{besteman04a,besteman05a}.

From a theoretical point of view, the standard Poisson-Boltzmann
theory is not able to predict charge inversion.  A number of theories
such as integral equations
\cite{lozada82a,gonzalez-tovar85a,greberg98a,kjellander01a}, field
theoretic approaches \cite{netz99b,moreira01b}, and others, see for
example the references in \cite{hidalgo96a,joensson01a}, have been put
forward to describe this effect.  Most popular amongst these are the
simple \emph{one-component plasma} (OCP) theories, which originated in
the Wigner-Crystal theory by Shklovskii \cite{perel99a,shklovskii99b},
and have some descendants by Levin \cite{levin02a} and a recent one by
Pianegonda \etal{} \cite{pianegonda05a}. The basic idea of these
theories is, that on the surface of the colloid, the multivalent ions
form a Wigner crystal or at least a \emph{strongly correlated liquid}.
Therefore, the charge can be overcompensated by the multivalent ions.
The rest of the system is considered in a Poisson-Boltzmann-like
fashion.  Although these theories are pretty simple, they give good
predictions in many cases.

Simulations have confirmed the phenomenon of overcharging
\cite{messina00b,messina01a,tanaka01a,tanaka02a,tanaka03a}, and they
are mostly consistent with the OCP theories in the strong-coupling
regime, which is close to the Wigner-crystal ($T=0$) limit. A notable
exception is the work by Messina \etal{} \cite{messina02d}, which shows
that overcharging is also possible in situations that are very far
from the strong coupling limit. Here, the overcharging was attributed
to mostly an entropy-driven effect.

Realistic and in particular biological systems, very often contain a
significant amount of ``normal'' monovalent salt.  Therefore it is
interesting to see what happens to the charge reversal when 1:1 salt
is added to the system.

Experimentally, with growing concentration of monovalent salt, the
inverted charge seems to decrease \cite{quesada-perez03a}.  In
contrast to that, most OCP theories predict a monotonous increase of
the inverted charge when more salt is added.  In an extension to
Shklovskii's original theory, Nguyen, Grosberg and Shklovskii
\cite{nguyen00a,grosberg02a} even predict so-called ``Giant
overcharging'' \cite{nguyen00a} for high concentrations of monovalent
salt, where the effective charge is larger than the bare (negative)
charge of the colloid.  Levin's theory \cite{levin02a} predicts a more
moderate growth, while Pianegonda, Barbosa and Levin
\cite{pianegonda05a} have a more detailed view: for low salt
concentrations, an increase for growing salt concentrations is
predicted, which reverses into a decrease for higher concentrations.

Not many simulations have been performed of that situation, however,
all of them seem to support the experimental finding, that the amount
of reversed charge decreases with increasing salt content.  Tanaka and
Grosberg have performed simulations of the electrophoresis of a
colloid in salt solution in a regime of very high surface charge
\cite{tanaka03a}.  Martin-Molina
\cite{quesada-perez05a,martin-molina06a} \etal{} simulate the
overcharging at a planar wall in salty environment under more
realistic conditions, which should be valid in the limit of large
colloids.  Diehl and Levin \cite{diehl06a} have performed Monte-Carlo
simulations of a spherical colloid, where the $\zeta$-potential is
measured.

This work complements these simulations by analyzing the case of a
spherical colloid in salty environment in the experimental parameter
regime, and comparing the results to the various assumptions made in
the OCP theories.

\section{System}
\label{system}

The model system we are considering is a single, spherical, highly
charged colloid in a solution of asymmetric 3:1 salt and additional
1:1 salt in a cubic box with periodic boundary conditions and a side
length of $L = 225.8\,\Ang$, which corresponds to a spherical cell with
a radius $R_0 = 140\,\Ang$.

The simulation box contains four particle types that are defined via
their diameter $d_i$ and their charge $q_i$:
\begin{itemize}
\item $1$ charged colloid ($Q_0 = -300\,e_0$, $d_0 = 100\,\Ang$)
\item $200$ (+3) counterions ($q_\mathrm{(+3)} = 3\,e_0$,
  $d_\mathrm{(+3)} = 4\,\Ang$)
\item $(300 + n_\mathrm{salt})$ (-1) coions ($q_\mathrm{(-1)} = -1\,e_0$,
  $d_\mathrm{(-1)} = 4\,\Ang$)
\item $n_\mathrm{salt}$ (+1) counterions ($q_\mathrm{(+1)} = 1\,e_0$,
  $d_\mathrm{(+1)} = 4\,\Ang$)
\end{itemize}
The colloidal surface charge density is $\sigma_s =
0.95\,\mathrm{e_0}\mathrm{nm}^{-2} =
0.152\,\mathrm{C}\mathrm{m}^{-2}$, which is in the experimentally
relevant regime.  $200$ trivalent counterions correspond to a
concentration of $c_\mathrm{(+3)}= 30\,\mathrm{mM}$.  Up to
$n_\mathrm{salt} = 1300$ monovalent salt ion pairs were added
(concentration $\csalt = 196\,\mathrm{mM}$).  The system is overall
electro-neutral and the colloid is fixed in the center of the
simulation box.

All particles interact via the Coulomb interaction 
\begin{equation}
  V_\mathrm{coulomb}(r) = \ell_b \frac{q_i q_j}{r}
\end{equation}
where the $q_i$ are the charges of the respective particle types and
$\ell_b = 7.1\,\Ang$ is the Bjerrum length in water at room
temperature.  Furthermore, the steric repulsion between the particles
is modelled via the core interaction
\begin{equation}
  V_\mathrm{core}(r) = \left\{
    \begin{array}{ll}
      4\varepsilon\left(
        \left(\frac{\sigma}{r-r_\mathrm{off}}\right)^{12}
        - \left(\frac{\sigma}{r-r_\mathrm{off}}\right)^{6}
      \right) & + V_\mathrm{shift}\\
      & \mathrm{,
        if~} r-r_\mathrm{off} < r_\mathrm{cut}\\
      0 & \mathrm{, otherwise} \\
    \end{array}
  \right. 
\end{equation}
This is the well-known shifted Lennard-Jones potential with an offset
of $r_\mathrm{off}$. To take into account only the repulsive part of
the potential, we choose $r_\mathrm{cut} = 2^{1/6}$ and
$V_\mathrm{shift} = \varepsilon$.  Also, we choose $\sigma = 1 \Ang$
and $\varepsilon = 1 \mathrm{kT}$ to ensure the same ``hardness'' of
all interactions, so that the distance of closest approach is
well-defined.  All particle sizes are modelled via the values of
$r_\mathrm{off}$ for the different interactions, which is defined by
$r_\mathrm{off} = \frac{1}{2}(d_i+d_j) - \sigma$, where $d_i$ and
$d_j$ are the diameters of the respective particle types.

The system was simulated using a standard molecular dynamics (MD)
algorithm using the Velocity Verlet algorithm in the (N, V,
T)-ensemble, with an MD time-step of $\delta t = 0.01$.  To speed up
the computation of the long-ranged electrostatic interactions, the P3M
algorithm \cite{hockney81a} was employed and tuned such, that the
maximal error in the forces was $\Delta_\mathrm{max} = 0.001$.  The
water environment was modelled implicitly via the Langevin thermostat
($T=1$, $\gamma=0.5$).  Hydrodynamic interactions were neglected,
as we are only interested in static equilibrium observables.

All simulations were performed using the simulation package ESPResSo
\cite{limbach06a}.


\section{Results}

\begin{figure}
  \centering
  \includegraphics[width=0.9\linewidth]{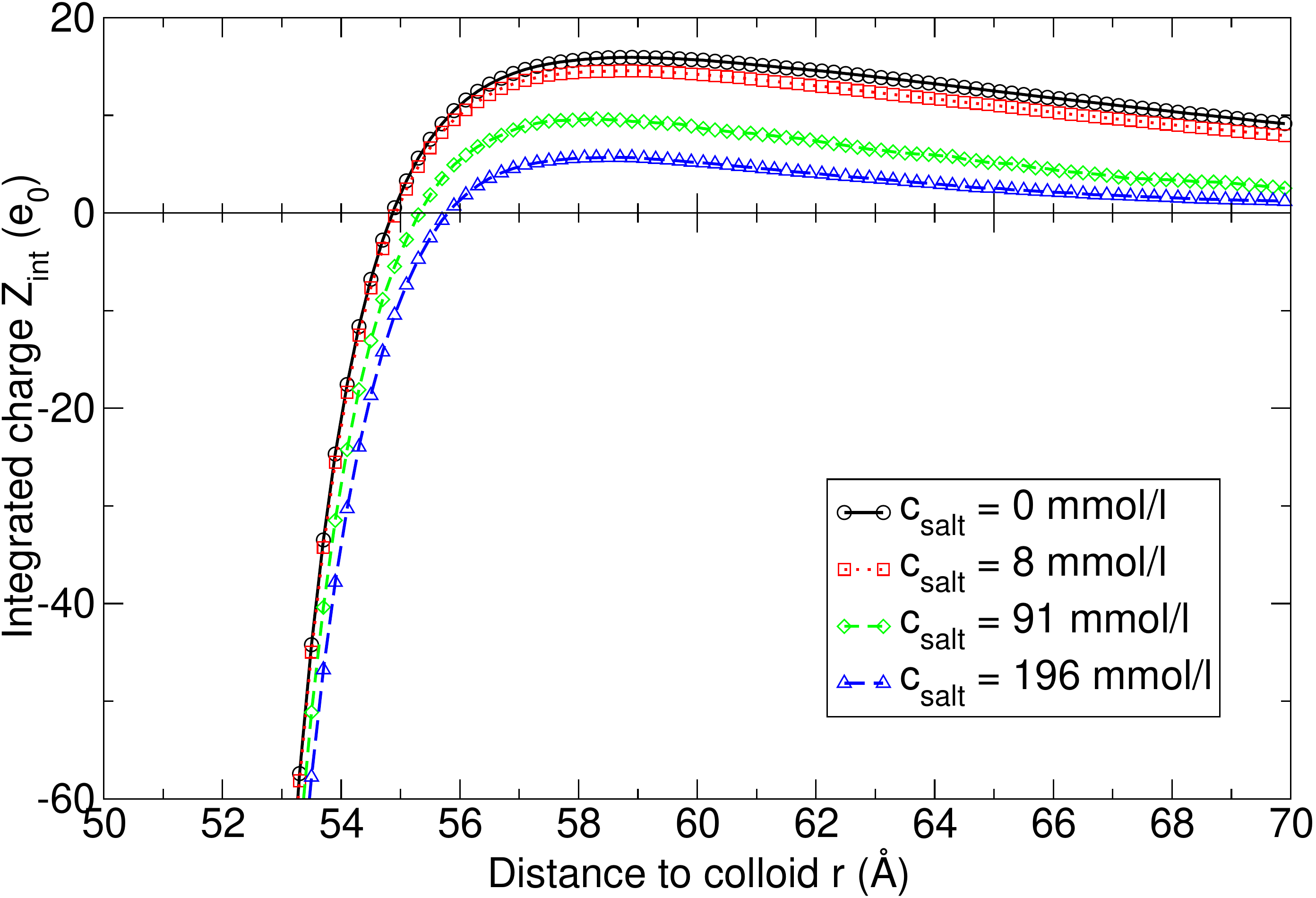}
  \caption{Integrated charge \zi{} against the distance from the
    colloid center $r$ for different concentrations of additional salt
    \csalt.}
  \label{fig:zi}
\end{figure}

Figure \ref{fig:zi} shows a plot of the integrated charge $\zi(r)$
against the distance to the center of the colloid for different
concentrations of the additional salt \csalt, which is defined by
\begin{equation}
  \zi(r) = Q_0 + \sum_i \int_0^r q_i \rho_i(r^\prime)\> dr^\prime
\end{equation}
where the $\rho_i(r)$ are the local densities of the different
particle types in a distance $r$ to the center of the colloid.  The
plot clearly depicts, that in all simulations, the (negative) charge
$Q_0$ of the colloid is overcompensated and has a maximum of up to $16
e_0$ at a distance of slightly more than two ion diameters from the
colloidal surface.

\begin{figure}
  \centering
  \includegraphics[width=0.9\linewidth]{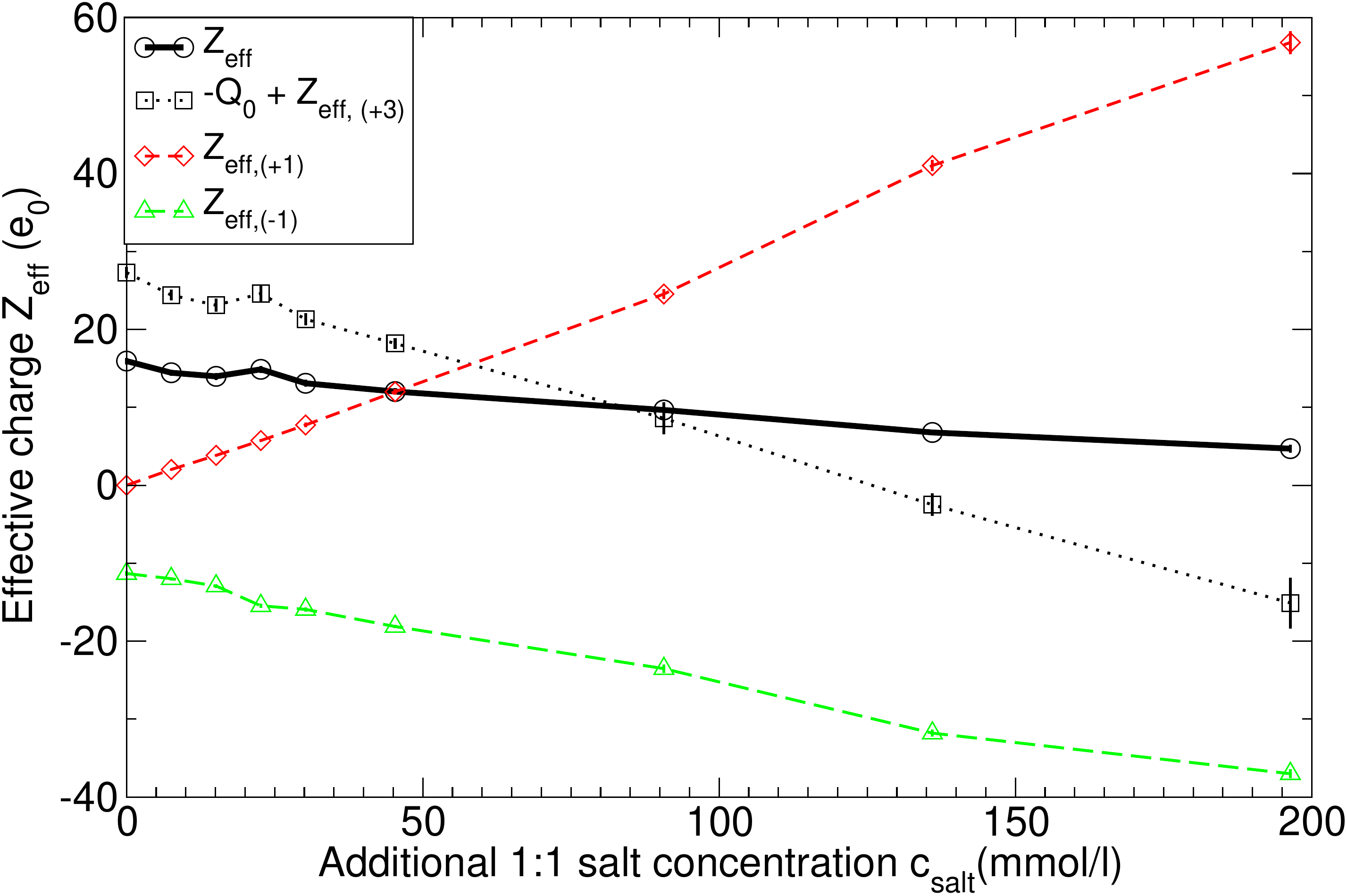}
  \caption{Effective charge \zeff (maximum of the integrated charge
    \zi) against the concentration of additional salt \csalt, and
    contributions of the different ion types.}
  \label{fig:zeff}
\end{figure}

In the following, we define the \emph{effective charge} \zeff{} of the
colloid as the maximum of the integrated charge distribution \zi.
This definition of the effective charge corresponds to the static
model of Tanaka and Grosberg \cite{tanaka01a,tanaka03a}.  

Figure \ref{fig:zeff} contains a plot of \zeff{} and the contributions
of the different ion types against the concentration \csalt{} of
additional monovalent salt.  Details are revealed by the plots of the
contributions of the different ion types in the figure.  Keep in mind,
that only by subtracting the bare charge $Q_0$ of the colloid from the
contribution of the multivalent (+3) ions $Z_\mathrm{eff, (+3)}$, the
different contributions are brought to the same scale.  This stresses,
that at all values of \csalt, the multivalent ions are responsible for
the largest part of the compensation of the colloid's bare charge.

Note, that our definition of \zeff{} is not identical to the
definition used in any of the previously cited theories, where only the multivalent
ions contribute to the effective charge.  To be able to directly
compare with the theories, one can look at the plot of the quantity
$-Q_0 + Z_\mathrm{eff, (+3)}$, \ie{} the contribution of the
multivalent ions $Z_\mathrm{eff, (+3)}$ minus the bare charge $Q_0$ of
the colloid.  In contrast to what Shklosvkii's and Levin's theories
predict, this quantity monotonously decreases with growing salt
concentration.  Also, it does not show a maximum, as the theory of
Pianegonda \etal{} would predict \cite{pianegonda05a}.

However, it is also apparent, that the number of multivalent ions that
contribute to \zeff{} decreases with growing concentration of
additional salt, up to a point where the multivalent ions alone would
not suffice to compensate $Q_0$ anymore.  This is in contradiction to
a fundamental assumption of the OCP theories, that we will discuss in
the following section.

Instead, the contribution of monovalent (+1) counterions
$Z_\mathrm{eff, (+1)}$ rapidly increases and is only partly
compensated by the also increasing (negative) contribution of
monovalent coions $Z_\mathrm{eff,(-1)}$.

\subsection{Strongly correlated liquid but no screening}
\label{sec:screening}

\begin{figure}
  \centering
  \includegraphics[width=0.9\linewidth]{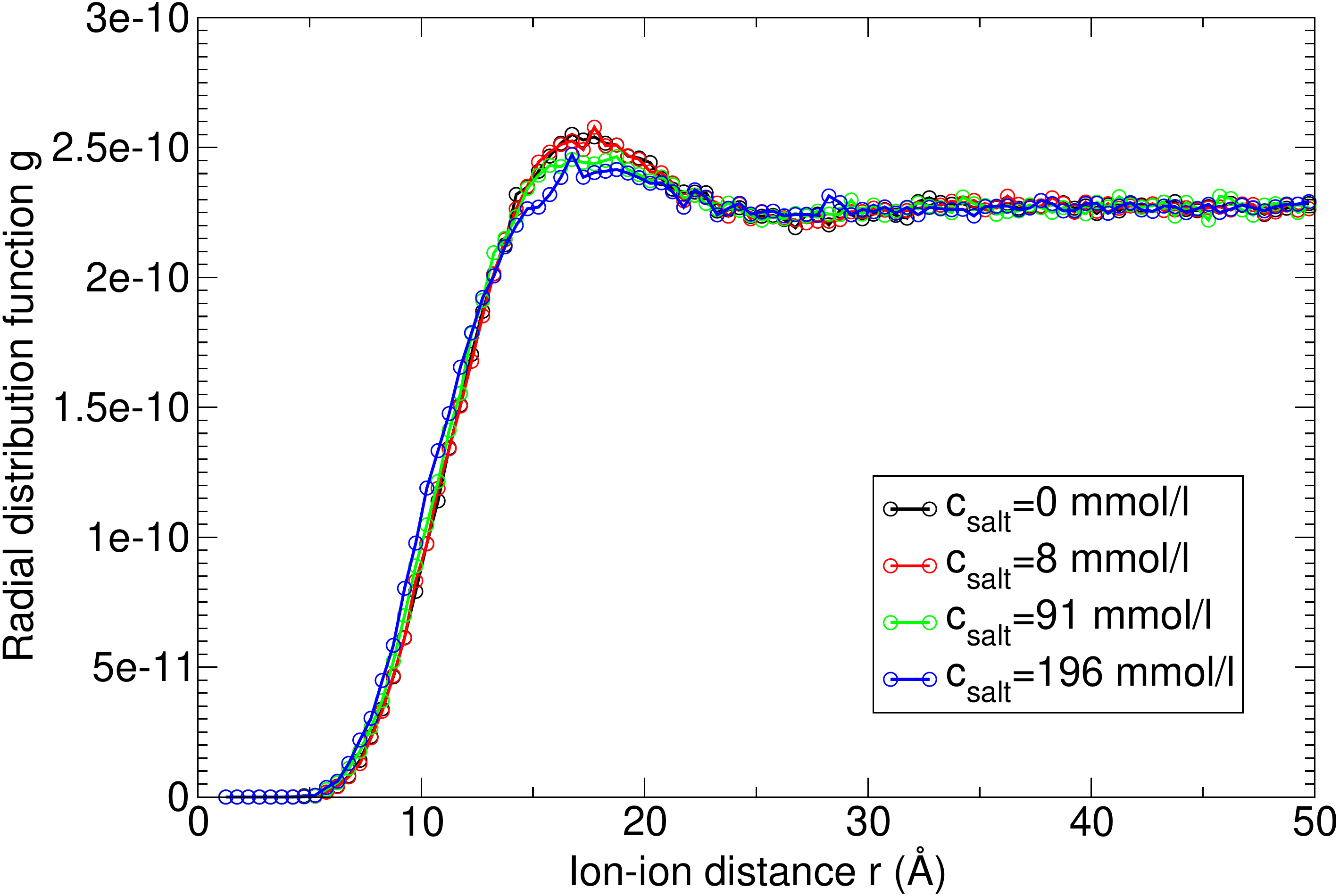}
  \caption{Radial distribution function $g(r)$ of the trivalent
    counterions on the colloidal surface that contribute to the
    effective charge for different concentrations of additional 1:1
    salt \csalt.}
  \label{fig:rdf2d}
\end{figure}

\begin{figure}
  \centering
  \begin{minipage}[t]{0.45\linewidth}
    \begin{center}
      \includegraphics[width=0.9\linewidth]{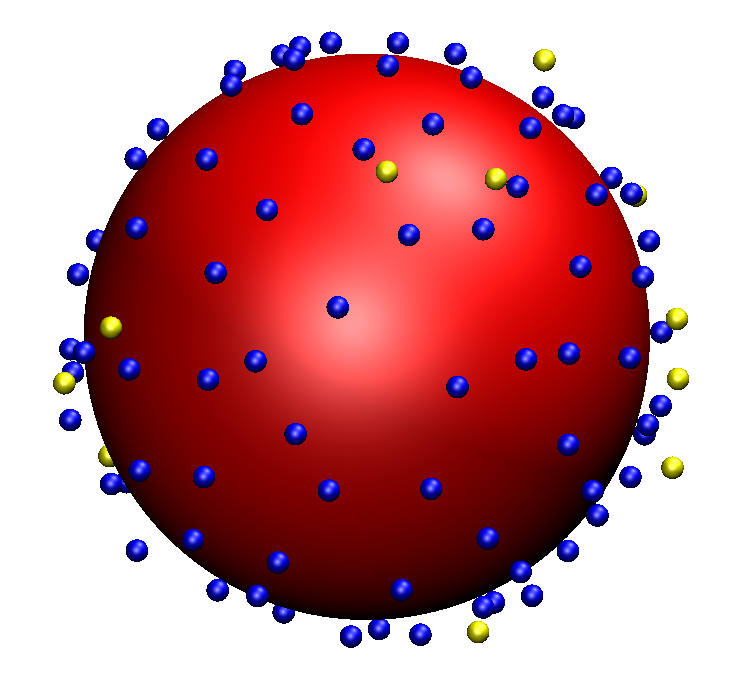}

      \small $\csalt = 0\,\mathrm{mmol/l}$
    \end{center}
  \end{minipage}

  \begin{minipage}[t]{0.45\linewidth}
    \begin{center}
      \includegraphics[width=0.9\linewidth]{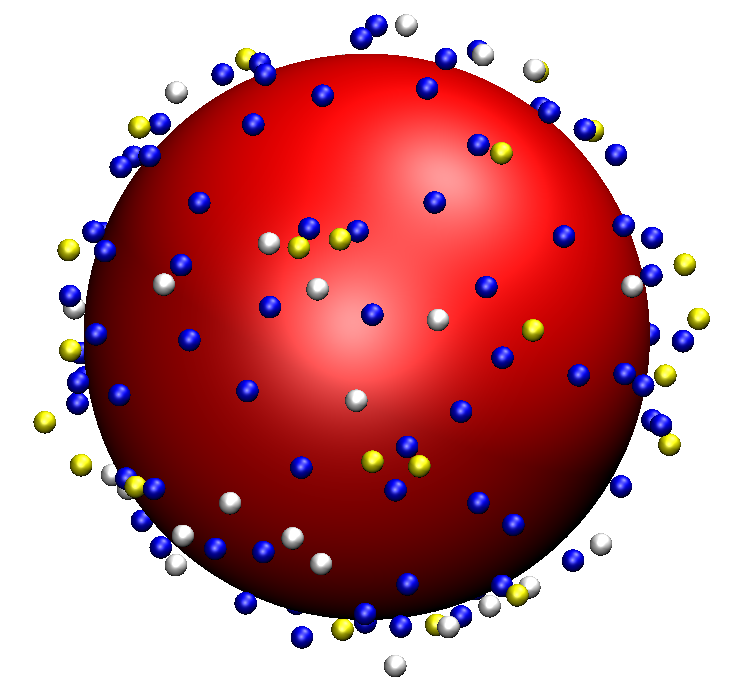}

      \small $\csalt = 91\,\mathrm{mmol/l}$
    \end{center}
  \end{minipage}
  \hfill
  \begin{minipage}[t]{0.45\linewidth}
    \begin{center}
      \includegraphics[width=0.9\linewidth]{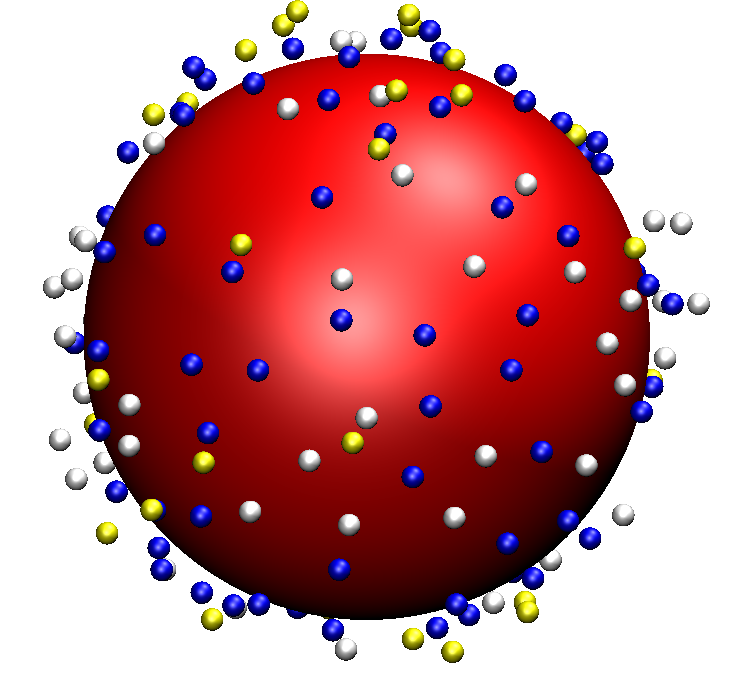}

      \small $\csalt = 196\,\mathrm{mmol/l}$
    \end{center}
  \end{minipage}

  \includegraphics[scale=0.3]{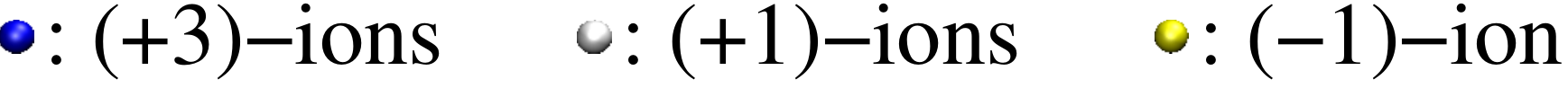}

  \caption{Snapshots of the colloid and the ions at the surface that
    contribute to the effective charge \zeff{} for different
    concentrations of additional 1:1 salt \csalt.}
  \label{fig:ss}
\end{figure}

All OCP theories predict, that the mutlivalent ions form a
\emph{strongly correlated liquid} on the colloidal surface, provided
that the coupling is large enough.  The plasma parameter of our
systems is $\Gamma_{2D} = (\pi \sigma_s \ell_B v^3)^{1/2} = 6.3$,
therefore we are in the strong coupling regime.  This view is
supported by the plot of the radial distribution function $g(r)$ of
the trivalent ions that contribute to \zeff{} in figure
\ref{fig:rdf2d}, which shows a correlation hole at very low distances
and a (weak) peak at a distance of about $a = 17\,\Ang$.  Also, the
simulation snapshots in figure \ref{fig:ss} show, that the trivalent
counterions indeed bind and completely cover the colloidal surface and
form what looks like a liquid-like structure.

According to Shklosvkii's theory, the only effect of adding monovalent
salt to the system should be to enhance the Debye-screening of the
electrostatic interactions between the multivalent ions.  This effect
should be most prominent, when the Debye screening length of the
monovalent salt, which is defined by
\begin{equation}
  r_S = (8\pi\ell_B\csalt)^{-1/2}
\end{equation}
is smaller than the preferred distance $a$ between the multivalent
ions.  In our simulations, we used values of the screening length of
down to $r_S = 8\,\Ang$, which is clearly smaller than the preferred
distance $a$ between the trivalent ions.  Still, $g(r)$ does not show
a shift in the position of the first peak, which would be expected, if
the preferred ion distance $a$ would be reduced.  This refutes the
idea, that Debye screening plays a significant role for the structure
of the multivalent ions close to the surface.

Instead, we see a small decrease in the height of the peak of $g(r)$,
meaning that the ions density decreases.  This observation is
consistent with the fact that less trivalent ions contribute to the
effective charge at higher salt concentrations.

\subsection{Monovalent ions on the surface}

\begin{figure}
  \centering
  \includegraphics[width=0.9\linewidth]{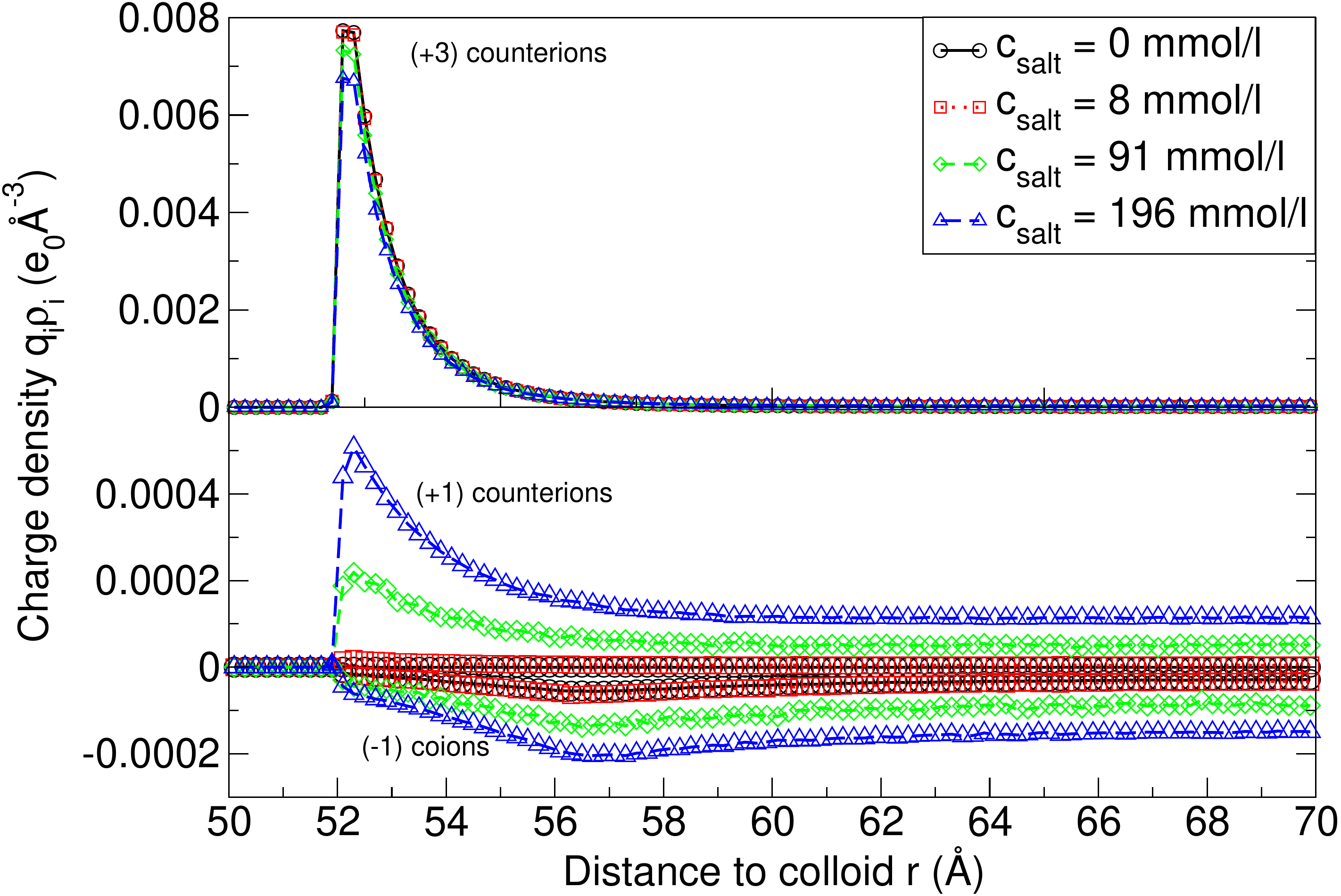}
  \caption{Contribution of the different ion types to the charge
    density $\rho$ against the distance from the colloid center $r$ for
    different concentrations of additional salt \csalt.  Note the
    different scales on the $y$-axis!}
  \label{fig:charge_density}
\end{figure}

As was shown above, even though the contribution of the trivalent ions
to the effective charge is most important, also the monovalent
counter- and coions contribute to the effective charge \zeff.  Where
are those ions located?

Figure \ref{fig:charge_density} shows a plot of the charge densities
$q_i \rho_i$ of the different ion types against the distance to the
colloid center $r$.  As was to be expected, with increasing \csalt,
the density of monovalent ions close to the colloid increases.
Interestingly, the maximum of the monovalent counterion density is at
the same distance to the colloid as the maximum of the trivalent
counterions for all salt concentrations.  Note, however, that the
maximum of the integrated charge \zi{} is at much larger values of $r
\approx 60\,\Ang$.

While the value at the maximum of the density of the (+1)-ions
increases with growing \csalt, the maximum value of the density of the
multivalent ions decreases, \ie{} some of the multivalent counterions
are replaced by monovalent counterions, which reduces the effective
charge.  This fact shows a weakness in the OCP theories, as these do
not take into account that monovalent counterions play any role for
the structure on the colloidal surface.

On the other hand, the monovalent coions form a weak, smeared-out
second layer, that can also be seen as an indicator of the charge
reversal of the colloid.

\subsection{Multivalent ions in the bulk}

\begin{figure}
  \centering
  \includegraphics[width=0.9\linewidth]{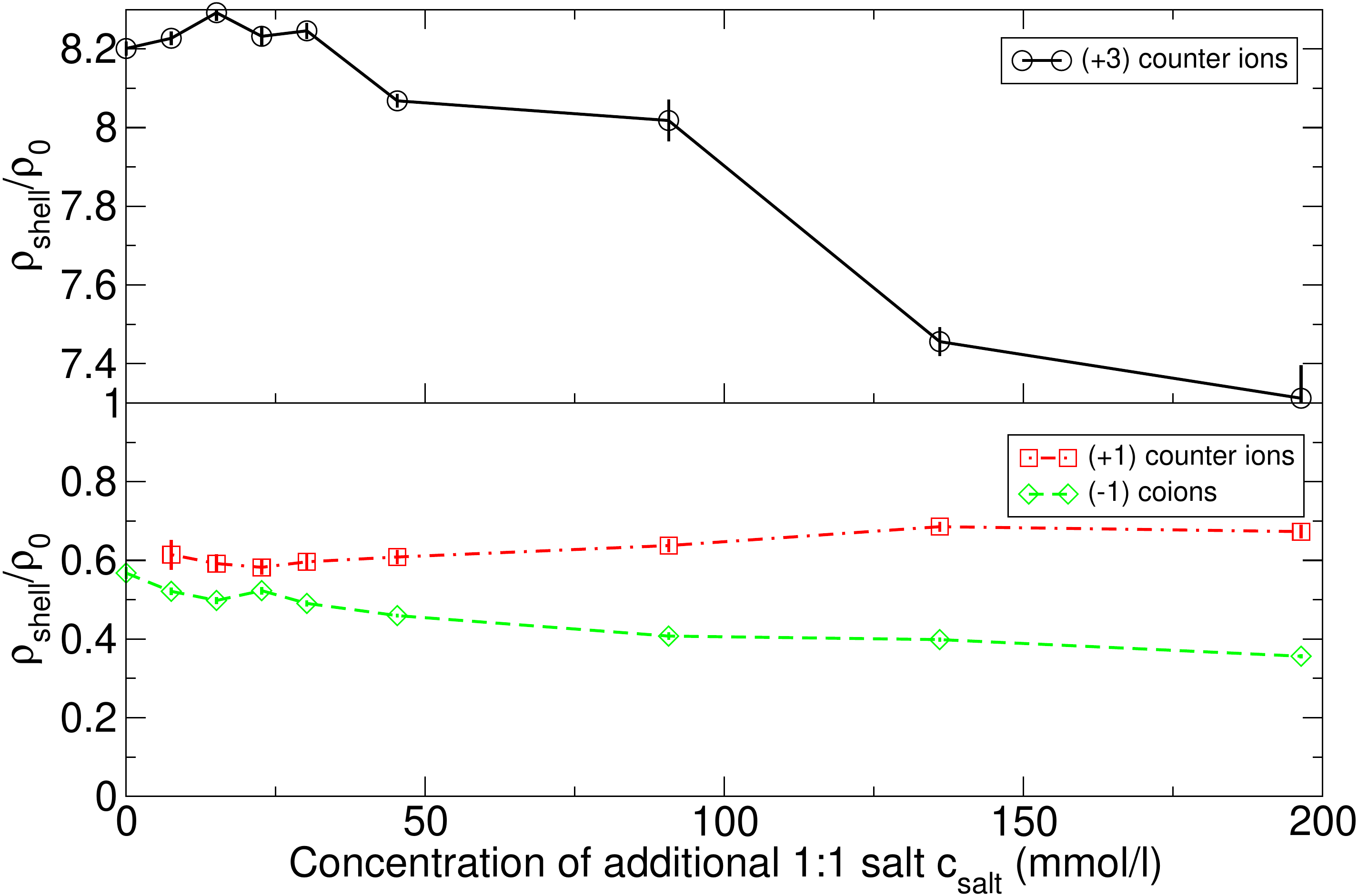}
  \caption{Densities of the different ion types on the colloidal
    surface relative to the corresponding bulk densities
    $\rho_\mathrm{shell,i}/\rho_{0,i}$ against the concentration of
    additional 1:1 salt \csalt.}
  \label{fig:rel_density}
\end{figure}

Figure \ref{fig:rel_density} shows a plot of the quantitity
$\rho_\mathrm{shell,i}/\rho_{0,i}$, where the $\rho_{0,i} = N_i/L^3$
are the bulk densities of the different ion types and the
$\rho_\mathrm{shell,i}$ are the respective densities of the ions in
the shell that contributes to the effective charge.  Not surprisingly,
at $\csalt = 0$, the density of the trivalent ions close to the
colloid is much greater than the bulk density of the trivalent ions,
while the density of the monovalent coions is smaller than in the
bulk.

The value of $\rho_\mathrm{shell,(+1)}/\rho_{0,(+1)}$ for the
monovalent counterions is almost constant for increasing salt
concentration \csalt, which indicates that the increasing contribution
of monovalent (+1) counterions to the effective charge is simply due
to the growing bulk density of these ions.

On the other hand, the value of $\rho_\mathrm{shell,i}/\rho_{0,i}$ of
the (+3)-counterions and the (-1)-coions decreases when more salt is
added.  This shows, that with growing salt concentration, more and
more of these ions prefer to stay in the bulk.

This can be interpreted in favour of the theory of Pianegonda \etal{}
\cite{pianegonda05a}, who take into account that the (negative)
solvation energy of the multivalent ions in the bulk is larger when
more salt is there.

This effect reduces the reversed charge with growing salt
concentration, which is also the reason why the theory of Pianegonda
\etal{} is the only OCP theory that correctly describes a decrease in
the effective charge at higher salt concentrations.

\section{Conclusions}

To conclude, our simulations give support to the basic assumption of
the OCP theories, namely that the overcompensation of the colloidal
charge by multivalent ions is well described by a strongly correlated
liquid.  However, the theories have a few shortcomings when it comes
to the description of the effect of additional monovalent salt.

In particular, the theories predict, that with growing salt
concentration, the strongly correlated liquid can pack more densly on
the colloidal surface, as the multivalent ions are screened by the
monovalent salt.  This assumption is not confirmed in our simulations.
High salt concentrations seem not to influence the size of the
Wigner-Seitz cell, and consequently, we do not see \emph{giant charge
  reversal}.  The giant overcharge observed in the parameter regime of
Tanaka and Grosberg \cite{tanaka01a} is in our point of view due to
the special choice of parameters.  For the strong coupling regime
considered, all ions form Bjerrum pairs, hence the concept of
screening becomes meaningless.

Instead, with growing salt concentration, monovalent counterions are
pushed onto the colloidal surface, and they replace some of the
multivalent ions, which lowers the effective charge.  This effect is
not included into the current OCP theories.

Finally, we observed, that not only the trivalent counterions are
pushed out of the region close to the surface of the colloid when more
salt is added, but also some of the monovalent coions leave the
region.  This is consistent with the assumption of Pianegonda \etal{}
\cite{pianegonda05a}, that the multivalent counterions and the
monovalent coions can gain favourite solvation energy when more salt
is in the bulk.

Therefore we conclude, that the theories of Shklovskii and Levin fail
to describe the phenomenon of colloidal charge reversal under the
influence of additional 1:1 salt correctly, as they are missing
important concepts.  On the other hand, the more recent theory of
Pianegonda \etal{} \cite{pianegonda05a} contains the necessary
ingredients for a good description, even though the quantitative
agreement with simulations is still not satisfactory
\cite{martin-molina06a}.


\begin{acknowledgement}
  The authors would like to thank Marcia C. Barbosa, Boris I.
  Shklovskii and Yan Levin for fruitful discussions, the PROBRAL
  project of the Deutscher Akademischer Austauschdienst (DAAD) for
  financial support of the exchange with M. Barbosa, and the Deutsche
  Forschungsgemeinschaft (DFG) for funding.
\end{acknowledgement}

\end{document}